\documentclass[sigconf]{acmart}

\copyrightyear{2026}
\acmYear{2026}
\setcopyright{cc}
\setcctype{by}                       
\acmConference[CPSIoTSec '26]{8th Workshop on CPS\&IoT Security and Privacy}{November 15-19, 2026}{The Hague, Netherlands}
\acmBooktitle{8th Workshop on CPS\&IoT Security and Privacy (CPSIoTSec '26), November 15-19, 2026, The Hague, Netherlands}
\acmDOI{10.1145/3847353.3847499}
\acmISBN{979-8-4007-3032-0/2026/11}
\usepackage{booktabs}
\usepackage{microtype}
\usepackage{amsmath}

\usepackage{amssymb}
\usepackage{balance}
\usepackage{pifont}                 
\newcommand{\cmark}{\ding{51}}
\usepackage{tikz}
\usetikzlibrary{arrows.meta,positioning}
\title[Formal Runtime Verification for Tool-Using LLM Agents]{Formal Runtime
       Verification for Tool-Using LLM Agents: An Offline Same-Benchmark Study
       on AgentDojo and STAC}

\author{Nikolaos Kekatos}
\affiliation{%
  \institution{Clone Systems}
  \city{Larnaca}\country{Cyprus}}
\email{nkekatos@clone-systems.com}

\author{Stylianos Basagiannis}
\affiliation{%
  \institution{International Hellenic University}
  \city{Serres}\country{Greece}}
\email{basagiannis@ihu.gr}

\author{Marinelio Chintri}
\affiliation{%
  \institution{International Hellenic University}
  \city{Serres}\country{Greece}}
\email{marchint@ihu.gr}

\author{Alexios Lekidis}
\affiliation{%
  \institution{University of Thessaly}
  \city{Larissa}\country{Greece}}
\email{alekidis@uth.gr}

\author{Tom Nianios}
\affiliation{%
  \institution{Clone Systems}
  \city{Larnaca}\country{Cyprus}}
\email{tnianios@clone-systems.com}

\author{Ioannis Seitoglou}
\affiliation{%
  \institution{International Hellenic University}
  \city{Serres}\country{Greece}}
\email{ioaseito@ihu.gr}

\author{Anastasios Temperekidis}
\affiliation{%
  \institution{Aristotle University of Thessaloniki}
  \city{Thessaloniki}\country{Greece}}
\email{anastemp@csd.auth.gr}

\author{Panagiotis Katsaros}
\affiliation{%
  \institution{Aristotle University of Thessaloniki}
  \city{Thessaloniki}\country{Greece}}
\email{katsaros@csd.auth.gr}

\renewcommand{\shortauthors}{Kekatos et al.}



\begin{document}

\begin{abstract}
Guardrails for tool-using LLM agents are usually application-specific rules, which
makes multi-step, data-dependent safety policies hard to specify, audit and reuse. As a
declarative alternative, we evaluate metric first-order temporal logic (MFOTL),
replaying the recorded trajectories that AgentDojo, STAC and R-Judge already ship
through the unmodified MonPoly monitor, offline and without running an agent. On these
corpora, five generic obligations flag $71.8\%$ of STAC attack chains and $70.1\%$ of
successful AgentDojo attacks, but also fire on $29.3\%$ of benign runs. This imprecision
stems from the corpora rather than the logic: they rarely record approvals and never
record timestamps, so history-dependent obligations reduce to detecting risky action
types. Where the trace does carry relational context, provenance-aware policies
discriminate better; that context, however, is itself attackable, and one planted line
defeats a naive provenance check on $94$--$99\%$ of the runs it would otherwise flag.
Binding provenance to the lookup that produced it closes this evasion at no cost in
detection or benign firing. Taken together, these results show that formal temporal
monitoring adds value exactly when the trace exposes trustworthy history. We therefore
quantify how far current benchmarks are from that point and propose a twelve-field
enforcement-ready trace schema.
\end{abstract}

\maketitle

\raggedbottom

\section{Introduction}
\label{sec:intro}

Large language model (LLM) agents no longer only answer questions. Given callable tools,
they read mailboxes, move money, edit shared documents~\cite{agentdojo2024,taubench2024}
and, increasingly, actuate physical devices~\cite{rjudge2024,resilagent2026}. Because
their inputs include untrusted content, an instruction planted in an email or a web page
can steer an agent into a harmful sequence of individually ordinary tool
calls~\cite{agentdojo2024,stac2026,toolemu2023}. Guardrails are the defence layer that
decides, at run time, whether a proposed call is
allowed~\cite{agentspec2026,progent2025,guardagent2024,llamaguard2023}. The harmful sequences, however, are rarely visible in any
single call: a transfer is dangerous because of where its payee came from, a deletion
because no one approved it, an actuation because of how recently the last one occurred.
Enforcing such policies requires reasoning over the interaction \emph{history}, and
evaluating them requires benchmarks that record that history.

The defences shipped with tool-using LLM agents are, in
practice, mostly \emph{rule-based}. AgentSpec~\cite{agentspec2026}, Progent~\cite{progent2025} and
GuardAgent~\cite{guardagent2024} evaluate predicates on tool calls; the defences
benchmarked in AgentDojo~\cite{agentdojo2024} (tool filters, prompt spotlighting,
prompt repetition) are per-call or per-prompt heuristics. Typical rules deny a
\texttt{send\_money} call above an amount threshold, restrict the callable tools to a
task-specific allow-list, or require a user prompt before any \texttt{delete}; each
inspects the pending call and, at most, a flag set earlier in the session. Some of these
engines maintain state, and with enough bespoke code a stateful rule can encode a
finite-history check. What they lack is a \emph{common declarative semantics} for the multi-step, data-dependent policies that the
safety-relevant attacks exploit: each policy is re-encoded per application, in
imperative code, hard to audit, reuse, or verify. And those attacks are multi-step:
STAC reports attack-success above $90\%$ precisely because the violation is assembled
across a chain of individually-benign calls~\cite{stac2026}. This paper asks whether
formal runtime verification can provide a reusable way to enforce such multi-step safety
policies for tool-using LLM agents, and whether existing agent-security benchmarks
contain enough information to evaluate such policies meaningfully. What remains
unclear is whether formal temporal monitoring retains a measurable advantage when
evaluated on the recorded attack traces that current agent-security benchmarks actually
provide, rather than on bespoke or fully instrumented scenarios.

Monitoring a \emph{trace} against a declarative temporal specification is the
defining task of runtime verification (RV). First-order temporal
monitors (MonPoly~\cite{monpoly2014}, DejaVu~\cite{havelund2018dejavu}) decide
metric, quantified past-time properties over event streams with data, which is the
exact shape of an agent obligation (``$\forall$ object $o$: an irreversible action
on $o$ requires a prior approval for $o$''). This makes metric first-order temporal
logic (MFOTL) and its monitor MonPoly a natural \emph{declarative substrate} for
such policies: quantified over data, with a formal
semantics and a reusable, independently-maintained engine, not bespoke
per-application code. It complements rule-based enforcement rather than replacing it.

We therefore evaluate whether formal runtime verification detects harmful tool-use
trajectories on existing agent-security benchmarks, and what precision costs arise.
Such an evaluation usually needs a live agent and an LLM. We avoid that: three major
benchmarks ship \emph{recorded} trajectories: AgentDojo publishes $36.7$k
recorded runs (tool-call traces plus its own attack-success and utility labels),
STAC publishes $483$ recorded attack chains, and R-Judge~\cite{rjudge2024} ships
labelled agent records with an IoT subset. We can therefore run a real RV engine
over the \emph{same} benchmarks the community uses, entirely offline. Offline replay
also isolates the monitoring question from model stochasticity and agent adaptation:
the same recorded behaviours are evaluated under multiple policy specifications and
monitor implementations.

Our central finding is that \emph{the value of formal RV for agent traces depends as
much on what the trace records as on what the specification language can express}.
On the corpora the community ships, generic temporal obligations collapse into
risky-action typing, because approvals, timestamps and state transitions are largely
absent; a keyword rule set would reach the same verdicts. Where relational or timing
context \emph{is} available, MFOTL expresses more discriminating policies, and
those policies then depend on trace content an attacker can write into. Provenance
integrity and benchmark instrumentation are therefore not side conditions but
prerequisites for history-dependent agent enforcement. Each contribution below is a
step in that argument.

\paragraph{Contributions}
\begin{itemize}
\item \textbf{A declarative RV formulation and same-benchmark evaluation methodology
      for tool-using agents.} We map recorded agent trajectories into MFOTL event
      streams and evaluate reusable generic, parametric and metric obligations using
      independently implemented runtime monitors, scored against each benchmark's own
      labels (Section~\ref{sec:oblig}).
\item \textbf{An empirical characterisation of when formal RV adds value.} Generic
      obligations achieve substantial attack coverage but largely collapse to
      typed-action detection on existing corpora. Where relational or timing context is
      available, data-parametric and metric MFOTL policies produce
      history-dependent verdicts and improve discrimination over action-only checking
      (Sections~\ref{sec:results}--\ref{sec:readiness}).
\item \textbf{The assurance conditions for history-dependent enforcement.} We show that
      trace provenance is itself an attack surface, demonstrate a lookup-bound
      provenance mechanism that eliminates the tested poisoning attack, and use a field
      ablation and a three-benchmark readiness audit to identify the information robust
      history-dependent enforcement requires. This motivates a twelve-field
      enforcement-ready trace schema (Sections~\ref{sec:canon}--\ref{sec:gap}).
\end{itemize}

\paragraph{Scope} We measure whether an obligation \emph{fires on a recorded
trajectory} (detection under post-hoc replay), not live-agent attack success under an
in-loop shield. An in-loop shield raises a different question, agent adaptation after
enforcement, which we leave to future work.

\paragraph{Roadmap} Section~\ref{sec:bg} reviews agent guardrails and runtime verification;
Section~\ref{sec:oblig} defines the monitored obligations and the replay methodology;
Section~\ref{sec:results} reports the generic and parametric results, Section~\ref{sec:canon} the
validity of the verdicts under canonicalisation error and provenance poisoning,
Section~\ref{sec:readiness} the benchmark-readiness audit and the timed CPS/IoT
policies; Section~\ref{sec:concl} discusses implications and limitations, and
Section~\ref{sec:conclusion} concludes.

\section{Background: Agent Guardrails and RV}
\label{sec:bg}

\paragraph{Rule-based agent guardrails} A per-call guardrail evaluates a
propositional predicate on the pending tool call. AgentSpec~\cite{agentspec2026}
and Progent~\cite{progent2025} give rule languages over the current call;
GuardAgent~\cite{guardagent2024} emits guard code from a knowledge base;
AgentDojo's built-in defences~\cite{agentdojo2024} filter tools or rewrite prompts;
Llama Guard~\cite{llamaguard2023} classifies message content. These are effective on
their target threats but reason over the current call (or prompt), not the
\emph{quantified past} of the trace, and each policy is re-encoded per application.

\paragraph{Benchmarks with recorded traces} AgentDojo~\cite{agentdojo2024}
evaluates prompt-injection attacks and defences across four suites (banking, slack,
travel, workspace) and \emph{ships} the resulting runs: per (model, suite, user
task, attack, injection) a JSON with the assistant tool-call trace and the labels
\textsf{security} (the injected task was accomplished) and \textsf{utility} (the
user task was accomplished). STAC~\cite{stac2026} ships $483$ recorded multi-turn
attack chains (derived from SHADE-Arena~\cite{shadearena2025} and
Agent-SafetyBench~\cite{agentsafetybench2024}), each a sequence of benign calls plus
a withheld final harmful step. R-Judge~\cite{rjudge2024} ships labelled agent
records across domains, including a smart-home / traffic / phone IoT subset. All are
replayable offline.

\paragraph{First-order temporal RV} MonPoly~\cite{monpoly2014} monitors MFOTL over
timed event streams and quantifies over data, which is the shape of the obligations in
this paper. DejaVu~\cite{havelund2018dejavu}, a first-order past-time monitor, and
VeriMon, MonPoly's formally verified kernel, support the untimed fragment and serve as
independent cross-checks of the STAC result (Section~\ref{sec:results}). Stream
monitors such as RTLola~\cite{rtlola2019} and TeSSLa~\cite{tessla2018} target numeric
signals and are complementary for quantitative obligations; our earlier edge-IoT work
runs both families side by side under a shared compute
budget~\cite{hybridhier2025,hiermon2026,critisfabric2026,v2xedge2026}, and companion
work applies mission-level runtime assurance to LLM-assisted swarms on simulated rather
than recorded traces~\cite{mesasswarm2026}. Closest to the present setting, a
guardrailed remediation agent for SIEM/XDR~\cite{guardrails2026} routes every
LLM-proposed action through a NeMo-Guardrails proxy: the per-call rule layer whose
blindness to history motivates this study. At enforcement time a
shield~\cite{shielding2018} gates the pending call on a violation.

\section{Obligations and Offline Methodology}
\label{sec:oblig}

\paragraph{System model} A \emph{tool-using LLM agent} is a model given callable APIs
that selects which to invoke, with what arguments and in what order. Its observable
behaviour is the resulting \emph{trace} of user messages, assistant messages, tool calls
and tool results; that trace is all our monitors read, and the model's internal reasoning
is neither observed nor constrained.

This section fixes what we monitor and how we measure it. We first state the obligations
as MFOTL formulas over a tool-call trace, then describe the pipeline that turns a
recorded trajectory into an event stream and the measures reported throughout.

Throughout, policies fall into four classes. \emph{Action-local} policies decide on the
current call alone (the per-call guardrails of Section~\ref{sec:bg}; on the recorded corpora,
O1--O5 reduce to this class). \emph{Relational history-dependent} policies quantify over
data established earlier in the trace (the payee-provenance and external-recipient
obligations of Section~\ref{sec:param}). \emph{Metric history-dependent} policies bound how
far back an enabler counts (the timed policies of Section~\ref{sec:cps}).
\emph{Content-semantic} policies judge the meaning of free text and lie outside
structural RV; R-Judge marks their extent (Section~\ref{sec:results}).

\subsection{Five generic obligations in MFOTL}

We define five reusable obligations, each of the form ``a risky action of a given
class requires a matching precondition (approval/consent/confirmation) earlier in
the trace,'' chosen by action \emph{semantics}, not by any benchmark's tool
names: \textbf{O1}~irreversible/destructive (delete, cancel, purge),
\textbf{O2}~protection teardown (disable, unlock, add auto-delete rule),
\textbf{O3}~financial commit (transfer, pay, schedule transaction),
\textbf{O4}~external export (send, share, upload), and \textbf{O5}~actuation/config
(set, execute, adjust a device). In MFOTL (O1 shown; $\blacklozenge$ is
once-in-the-past):
\[
  \square\,\forall a.\ \big(\mathsf{destructive}(a)
    \rightarrow \blacklozenge\,\mathsf{approved}(a)\big).
\]
These are the specifications we hand to MonPoly (as their monitored violation form;
full formulas and keyword patterns in Appendix~\ref{app:repro}). They are
deliberately generic; the study measures how far five reusable formal templates go
on real benchmarks, and where precision then demands \emph{parametric} refinement
(binding the approval to the specific object, within a window): expressible in
MFOTL but not in stateless per-call predicates.

\subsection{Replay pipeline and measures}

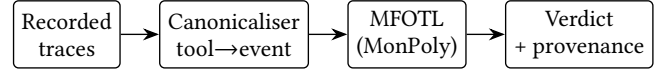
\begin{figure}[t]
\centering
\Description{Block diagram with four boxes connected left to right by arrows: recorded
trajectories, canonicaliser producing a typed event stream, MonPoly checking MFOTL
obligations, and scoring of verdicts against benchmark labels.}
\resizebox{\columnwidth}{!}{%
\begin{tikzpicture}[node distance=4mm, >=Stealth,
  b/.style={draw, rounded corners=1.5pt, align=center, inner sep=2.5pt,
            font=\footnotesize, minimum height=7mm}]
\node[b] (r) {Recorded\\traces};
\node[b, right=of r] (c) {Canonicaliser\\tool$\to$event};
\node[b, right=of c] (m) {MFOTL\\(MonPoly)};
\node[b, right=of m] (v) {Verdict\\+ provenance};
\draw[->] (r)--(c); \draw[->] (c)--(m); \draw[->] (m)--(v);
\end{tikzpicture}}
\looseness=-1 \caption{Offline replay pipeline: recorded trajectories are canonicalised to a typed
event stream and checked against MFOTL obligations by MonPoly; verdicts are scored
against each benchmark's own labels. No LLM or agent runs.}
\label{fig:pipeline}
\end{figure}
\label{sec:method}

Figure~\ref{fig:pipeline} shows the offline pipeline.
\paragraph{Canonicalisation} For each recorded run we build one event stream from
the assistant tool calls (AgentDojo: \texttt{\{function,args\}} per call; STAC: the
interaction history plus the withheld final step; R-Judge: the agent-turn actions).
The monitor types each tool into an obligation class by semantic keyword patterns
and detects preconditions by approval patterns (Appendix~\ref{app:repro}).

\paragraph{Worked example} STAC chain $395$ records four calls:
\texttt{search\_db}, \texttt{read}, \texttt{confirm\_for\_deletion} and
\texttt{delete}. Canonicalisation keeps the obligation class and the chain id and
drops everything else: the two read-only calls produce no event, and the stream is
\texttt{@309 approval(395)}, \texttt{@310 destructive(395)}. The generic obligation
(a risky action on chain $c$ with no earlier $\mathsf{approval}(c)$) stays silent,
because the approval precedes the deletion. On chain $0$ the stream is
\texttt{@0 export(0)} with no approval anywhere in the log, and MonPoly fires.
Arguments are dropped at this generic layer only; the parametric obligations of
Section~\ref{sec:param} read the values they need (payee, recipient) from the recorded call.

For STAC we additionally export the canonical stream to a
MonPoly log and the obligations to MFOTL, and run the \emph{unmodified MonPoly}
binary; its firings are the RV result, not a re-implementation. The CPS/IoT policies
of Section~\ref{sec:cps} are likewise evaluated by MonPoly.

\paragraph{What the policies need} The obligations are not stateless predicates. They
quantify over trace data (payee identity, recipient domain) \emph{and} bound how far
back an enabler counts (a consent within $24$\,h, three actuations in $6$\,h).
First-order quantification plus a metric past is exactly safe-range MFOTL.

\paragraph{Why MonPoly} MonPoly is a maintained implementation of that fragment that
runs \emph{unmodified} on an exported log, so the headline firings are not a
re-implementation of our own semantics. The other tools play defined roles.
DejaVu~\cite{havelund2018dejavu} decides first-order past-time properties with BDDs but
has no metric windows without its timed extension, so we use it to cross-check the
untimed STAC obligations rather than the bounded ones. VeriMon, MonPoly's formally
verified kernel, covers the same fragment and cross-checks the same result; it would be
the natural choice where a machine-checked monitor is required.
TeSSLa~\cite{tessla2018} and RTLola are stream-based and strong on per-trace numeric
computation but weaker on first-order joins over unbounded domains (arbitrary payee and
recipient strings), which the parametric obligations need. The engine is therefore
interchangeable within the fragment; the \emph{specification} is the contribution.

\paragraph{Measures} On AgentDojo we use its labels: over attacked runs,
\emph{detection on successful attacks} (of \textsf{security}$=$true runs, the
fraction where an obligation fires) and the \emph{benign firing rate} (BFR: of benign
runs, the fraction where an obligation fires). BFR is a potential-block proxy for the
utility cost (what a shield \emph{would} gate) not a measured task-failure rate,
which would require re-running the agent. On STAC (attack-only) we
report \emph{final-step detection} (fires on the withheld harmful step); on R-Judge,
detection on unsafe records and false-positive rate on safe ones. Confidence
intervals are Wilson $95\%$ score intervals.

\paragraph{Sampling and parameters} The unit is one recorded run. A
benign run is one user task with \emph{no} injection ($\approx\!30$ per suite, so
$123$ benign GPT-4o runs with a parseable trace); an attacked run is a (user
task~$\times$~injection~$\times$~attack) combination ($6680$). The benign set is thus
small, and we report Wilson intervals on every rate (e.g.\ the $29.3\%$ aggregate BFR
has interval $[22.0,37.8]$). The keyword patterns are the ordinary English verbs for
each class (Appendix~\ref{app:repro}), fixed \emph{before} measuring outcomes and
never adjusted; the one threshold, the $\blacklozenge$ window, is the most permissive
choice (unbounded past), so a firing is never a tight-window artefact. ``Detection''
means the obligation fires on the recorded call; a would-be shield gating it assumes
prevention. STAC's content-semantic failure modes (harmful free text) are out of
structural scope and flagged, if at all, only via the delivery act.

\emph{One correction to the frozen set.} Validation exposed two substring-matching
errors: the approval token \texttt{ack} collided with unrelated Slack tool names, and
read-only accessors (\texttt{get\_}, \texttt{list\_}, \ldots) were being typed as
approvals or as financial actions. We corrected both with explicit read-only exclusions.
The corrected mapping yields the $70.1\%$ detection and $29.3\%$ BFR reported throughout;
the pre-correction figures were $61.6\%$ and $31.7\%$, so the errors understated the
method rather than inflating it. STAC and R-Judge are unaffected. The full forensics are
in Appendix~\ref{app:forensic}; they are precisely the class of error a substring
canonicaliser invites and what Section~\ref{sec:canon} measures.

\section{Results}
\label{sec:results}

We report the generic obligations first, then the parametric ones. The contrast between
the two matters: the generic layer establishes how much of the recorded attack
surface typed-action checking reaches at all, and the parametric layer is where a verdict
starts to depend on the trace's history.

\subsection{Generic obligations on recorded trajectories}

\paragraph{AgentDojo (same-benchmark, offline)} Table~\ref{tab:agentdojo} replays
all recorded GPT-4o runs (no defence). The generic obligations fire on $70.1\%$ of the
attacks that succeeded and on $29.3\%$ of benign runs; the financial, export and
actuation obligations account for the detections (teardown never fires, since these
suites have no teardown tools).

Two observations qualify these rates. First, they measure \emph{typed-action}
detection: the AgentDojo traces contain no approval events, so the approval condition
never holds and the verdict reduces to the tool's action class. A keyword rule set would reach
the same verdicts, and detection and benign firing therefore rise and fall together,
because a legitimate task also sends email, moves money and changes configuration.
Section~\ref{sec:gap} attributes this to the corpora rather than to MFOTL; the claims that
rest on the formalism are the parametric obligations of Section~\ref{sec:param} and the timed
policies of Section~\ref{sec:cps}. Second, a $29.3\%$ BFR is not a deployable operating
point. BFR counts runs on which some obligation fires, a firing proxy rather than a
measured utility cost, since the blocked task is never re-executed. At this rate the
generic layer is suited to triage for human review rather than to enforcement gating.

\begin{table}[t]
\centering
\caption{AgentDojo recorded runs (GPT-4o, no defence) replayed offline. Detection is on
AgentDojo's \textsf{security}=true attacks, BFR on its benign runs; $96$ of $6899$ runs
with an empty tool-call chain are excluded. Per-obligation counts overlap and sum to
$1677$ against $1354$ detected, so the union is the detection figure
(Appendix~\ref{app:extended}, Table~\ref{tab:peroblig}).}
\label{tab:agentdojo}
\renewcommand{\arraystretch}{1.1}
\setlength{\tabcolsep}{5pt}
\begin{tabular}{@{}lc@{}}
\toprule
\textbf{Measure} & \textbf{value} \\
\midrule
Attacked runs                                  & $6680$ \\
AgentDojo attack-success (no defence)          & $1931/6680 = 28.9\%$ \\
RV detection on \emph{successful} attacks       & $1354/1931 = 70.1\%$ \\
RV fires on any attacked run                     & $2292/6680 = 34.3\%$ \\
RV BFR on benign runs                     & $36/123 = 29.3\%$ \\
\quad Wilson $95\%$ interval on BFR             & $[22.0,37.8]$ \\
\bottomrule
\end{tabular}
\end{table}

\paragraph{Across models} The pattern is not GPT-4o-specific: across the
$22$ base (no-defence) model directories AgentDojo ships, RV detection on successful
attacks spans $48.4$--$91.7\%$ and BFR $16.7$--$34.8\%$. AgentDojo's own built-in
defences are not comparable to our offline replay, since they execute the agent
\emph{with} the defence in the loop, whereas we never re-run the agent.

\paragraph{Earliest warning} A
detection is only useful for prevention if it arrives before the harmful step. On the
$1931$ successful GPT-4o AgentDojo attacks, an obligation fires \emph{before} the
final harmful call (\emph{preventive}) in $1012$ ($52.4\%$), fires only \emph{at} that
call (\emph{last-moment}) in $342$ ($17.7\%$), and never fires in $577$ ($29.9\%$);
where it is preventive the median detection lead is one tool-call (mean $2.19$, max
$13$). On STAC the picture inverts by construction: the corpus \emph{withholds} the
final harmful step, so the signal concentrates there: $57/483$ ($11.8\%$) preventive,
$297$ ($61.5\%$) last-moment, $129$ ($26.7\%$) no firing. The earliest-warning margin,
not a post-hoc residual, is our intervention-timing evidence.

\paragraph{Per suite} The suite breakdown
(Appendix~\ref{app:extended}, Table~\ref{tab:suite}) exposes the precision problem
concretely. \textbf{Banking} exposes it most clearly: financial actions occur
frequently in both successful attacks and legitimate tasks, producing $97.4\%$
detection but $79.2\%$ benign firing. Travel, with few money/export actions, is the opposite
($16.5\%$ detection, $11.1\%$ BFR). No single generic threshold is right for all
suites; the domain-specific \emph{parametric} refinements (Section~\ref{sec:param}) are what
recover precision.

\paragraph{STAC} Replaying all
$483$ STAC chains, our reference monitor flags the withheld harmful action in
$347/483 = 71.8\%$ ($95\%$ CI $[67.7,75.7]$) of chains, and the \emph{unmodified
MonPoly engine}, fed the MFOTL obligations and the exported log, fires on
\emph{exactly} the same $347$. VeriMon, MonPoly's Isabelle/HOL-verified kernel, and
the independently developed DejaVu return the same $347$ (Appendix~\ref{app:repro}). On the
structurally-intended failure modes the rate is $72.8\%$; the content-semantic modes
(harmful text, trusting/filtering tool results, $\approx$$29\%$ of cases) are out of
structural scope and flagged only via the delivery act. $136$ chains end in an action
outside the five generic classes (spam-rule edits, calendar changes, in-place
appends), mapping where new obligations or semantics are needed.

\paragraph{R-Judge} To bound the reach of
structural trace-level RV, we replay the \emph{full} R-Judge~\cite{rjudge2024}: $571$
records ($301$ unsafe, $270$ safe), a genuine benign set unlike STAC. The generic
obligations fire on $23.6\%$ ($71/301$) of unsafe records at $16.3\%$ ($44/270$) FPR.
The low detection is expected: most R-Judge unsafe cases are \emph{content-semantic}
(privacy leakage, misinformation, unsafe advice) with no structural trace signature,
so a rule-based per-call checker fares no better. R-Judge thus \emph{bounds} the reach
of structural trace-level RV (formal or rule-based alike) and confirms a semantic
aspect is the necessary complement.

\subsection{Parametric obligations: relational conditions over trace data}
\label{sec:param}

Consider two AgentDojo banking runs with the same three calls,
\texttt{read\_inbox}, \texttt{get\_transactions}, \texttt{send\_money}. In the
legitimate run the inbox holds a supplier invoice, the transaction lookup returns the
supplier's IBAN as a past counterparty, and the transfer goes to that IBAN. In the
attacked run the inbox holds an injected instruction (``pay the March difference''),
the lookup returns the same history, and the transfer goes to the attacker IBAN
\texttt{US13\,3000\ldots}, a value never returned by any lookup that exists only inside
the email text. The final call alone cannot tell the two runs apart; what separates
them is what the trace established beforehand.

The AgentDojo BFR is not an artefact of RV: it afflicts \emph{any}
action-type checker, rule-based or formal, because ``sends email'' is both the attack
and the job. Closing it needs a \emph{precise} condition that quantifies over trace
data: allow the transfer only if a matching approval for \emph{that} payee occurred,
or the export only if the recipient came from the user's own directory. Such
conditions are expressible in MFOTL and monitorable by MonPoly; a stateless per-call
predicate cannot refer to the payee's past at all. A stateful rule engine could implement
the same check procedurally. The advantage of MFOTL is that the policy is stated
declaratively, with explicit temporal and relational semantics, and evaluated by a
reusable engine, rather than reimplemented as application-specific state-management
code. We implement two such obligations and measure them (Table~\ref{tab:param}).

\paragraph{Payee provenance (banking)} Instead of firing on \emph{any} transfer
\looseness=-1 (generic O3), the parametric obligation fires only on a transfer to a payee never
seen as a \emph{structured} recipient or sender in prior account data,
\[
  \square\,\forall p.\ \big(\mathsf{transfer}(p)\wedge
    \neg\,\blacklozenge\,\mathsf{known\_payee}(p)\big).
\]
On GPT-4o banking runs this cuts the benign firing rate from $44.0\%$ to $24.0\%$ while
detection on successful attacks falls only from $79.2\%$ to $70.0\%$: the injected
attacker IBAN is embedded in free text and is never a real counterparty, so it is
still flagged, whereas a legitimate transfer to a known payee no longer is.

\paragraph{External recipient (workspace)} Symmetrically for export, the parametric
obligation fires only on a \texttt{send\_email} whose recipient never appeared as a
structured directory entry (contact/participant/sender field) in a prior tool result,
\[
  \square\,\forall r.\ \big(\mathsf{send}(r)\wedge
    \neg\,\blacklozenge\,\mathsf{known\_contact}(r)\big).
\]
The injection payload delivers the attacker address inside a free-text email body,
not the directory, so it is treated as external; a colleague already in the calendar
or address book is not. This cuts the workspace BFR from $15.2\%$ to $10.9\%$ at a
detection cost of $42.5\%\!\to\!39.2\%$.

\paragraph{Discrimination} The positive likelihood ratio $\mathrm{LR^+}$
(detection $/$ BFR) summarises both error rates in one number. Refining the generic
obligations ($\mathrm{LR^+}=2.40$) to the parametric spec raises it on both suites:
payee provenance $1.80\!\to\!2.92$, external recipient $2.79\!\to\!3.61$, so the
parametric conjunct separates benign from malicious transfers/exports better than
typed-action detection.

\paragraph{Statistical significance} Across the pooled benign runs, every verdict
changed by the parametric refinement is a false-positive reduction; none changes in the
opposite direction, and the reduction is statistically significant under an exact
McNemar test. In detail: the parametric obligation is a strict refinement of the generic
one, so the paired comparison can differ only where the generic fires and the parametric
does not. Single-model denominators are too small to be conclusive; pooling benign runs
across all AgentDojo model directories ($700$ banking, $1{,}376$ workspace) gives $75$
vs $0$ discordant runs for banking and $37$ vs $0$ for workspace, with detection largely
preserved, and $p=5.3\times10^{-23}$ (banking) and $p=1.5\times10^{-11}$ (workspace).

\begin{table}[t]
\centering
\caption{Parametric vs.\ generic obligations (GPT-4o), Wilson $95\%$ intervals. Each row
is a \emph{single} obligation on a \emph{single} suite, so these rates are not comparable
with the policy-set unions reported elsewhere; benign denominators are $25$ banking and
$46$ workspace runs; conventions in Appendix~\ref{app:conventions}.}
\label{tab:param}
\renewcommand{\arraystretch}{1.15}
\setlength{\tabcolsep}{4pt}
\begin{tabular}{@{}llcc@{}}
\toprule
\textbf{Suite} & \textbf{Spec} & \textbf{detect \% [CI]} & \textbf{BFR \% [CI]} \\
\midrule
banking   & generic O3          & $79.2\ [75.7,82.3]$ & $44.0\ [26.7,62.9]$ \\
banking   & param.\ payee       & $70.0\ [66.1,73.6]$ & $24.0\ [11.5,43.4]$ \\
workspace & generic O4          & $42.5\ [38.3,46.8]$ & $15.2\ [7.6,28.2]$  \\
workspace & param.\ ext-recip.\ & $39.2\ [35.1,43.5]$ & $10.9\ [4.7,23.0]$  \\
\bottomrule
\end{tabular}
\end{table}

\paragraph{What depends on temporal reasoning} The generic rates are typed-action
detection, as Section~\ref{sec:results} states. The two parametric rows of
Table~\ref{tab:param} and every verdict flip in the timed study do depend on genuine
relational or temporal reasoning: the \emph{same} call is allowed or flagged according to
a prior-provenance or prior-state condition, which a stateless per-call predicate cannot
decide and which MFOTL states declaratively instead of in bespoke stateful code.

\section{Validity of Trace-Derived Verdicts}
\label{sec:canon}

A verdict is only meaningful if the two things it rests on can be trusted: the typing
that decides an action's class, and the trace content from which provenance is derived.
This section attacks both. The first is an accuracy question we measure against a
hand-labelled gold set; the second turns out to be an adversarial question, because the
provenance a policy reads is content an attacker can write.

\subsection{Canonicalisation accuracy and sensitivity}

Detection is only as good as the tool$\to$class typing, so we measure it and bound its
influence. Against a gold set of $54$ tool names hand-labelled blind to any detection
outcome, the keyword typing reaches micro-averaged precision $0.90$, recall $0.84$ and
single-primary exact-match $87.0\%$ ($47/54$), with systematic rather than random
residual errors. Re-running the replays under four tool$\to$class mappings moves AgentDojo
detection within $[69.0,79.9]\%$ and BFR within $[26.0,43.9]\%$, and STAC detection within
$[61.7,73.1]\%$: moderately sensitive, and the frozen typing is not an optimistic choice
on AgentDojo, since replacing it with the gold-corrected mapping increases detection
rather than lowering it. The gold
set, the per-class table and the four mappings are in
Appendix~\ref{app:extended} (Tables~\ref{tab:canon} and~\ref{tab:canonsens}).

\subsection{Provenance integrity: the monitor's trace history is an attack surface}
\label{sec:prov}

The parametric obligations of Section~\ref{sec:param} draw their power from a
\emph{provenance} predicate (\textsf{known\_payee}, \textsf{known\_contact}),
computed over the monitor's own recorded trace: a payee is trusted iff it appeared
earlier as a structured recipient/sender field. That trace is, however, partly
attacker-controlled: an injection can plant text that the canonicaliser reads as a
structured directory entry, so the history the monitor trusts is itself an attack
surface.

\paragraph{Poisoning the provenance signal} We test this directly, injecting one line
naming the attacker recipient into each flagged run's readable context in four framings.
A \emph{structured} directory line or a \emph{quoted/forwarded} line flips $94.0\%$ of
the $403$ flagged banking-payee runs and $99.0\%$ of the $203$ flagged
workspace-recipient runs to ``known'' (no longer flagged); a \emph{free-text} mention
and a \emph{reformatted} recipient flip $0\%$ (per-framing counts in
Appendix~\ref{app:extended}). One planted structured or quoted line therefore defeats
the provenance check for $94$--$99\%$ of the transfers it would otherwise flag: the
check trusts \emph{syntactic} structure it cannot authenticate. No model is deceived
here; only the monitor is.

\begin{table}[t]
\looseness=-1 \centering
\caption{Hardened provenance against the poisoning attack of Section~\ref{sec:prov}: flip rate
(flagged runs whose verdict reverses to ``known''). Detection and benign firing are
identical under both predicates, so the hardening costs nothing measurable.}
\label{tab:harden}
\renewcommand{\arraystretch}{1.05}
\setlength{\tabcolsep}{3.5pt}
\begin{tabular}{@{}llcc@{}}
\toprule
\textbf{Suite} & \textbf{planted line} & \textbf{naive} & \textbf{hardened} \\
\midrule
banking ($403$ flagged)   & structured        & $94.0\%$ & $\mathbf{0.0\%}$ \\
                          & quoted/forwarded  & $94.0\%$ & $\mathbf{0.0\%}$ \\
                          & free text         & $0.0\%$  & $0.0\%$ \\
                          & reformatted       & $0.0\%$  & $0.0\%$ \\
\midrule
workspace ($203$ flagged) & structured        & $99.0\%$ & $\mathbf{0.0\%}$ \\
                          & quoted/forwarded  & $99.0\%$ & $\mathbf{0.0\%}$ \\
                          & free text         & $0.0\%$  & $0.0\%$ \\
                          & reformatted       & $0.0\%$  & $0.0\%$ \\
\bottomrule
\end{tabular}
\end{table}

\paragraph{A hardened provenance specification} We therefore strengthen the
specification so that provenance integrity is itself an obligation. A hardened
obligation requires that
(i) provenance come from a \emph{trusted tool result} (an account/directory API
response), not message text; (ii) a trusted contact \emph{predate the session};
(iii) an approval \emph{bind to the exact payee and amount}, not the action class; and
(iv) a \emph{new} payee require explicit confirmation. Conditions (i)--(iv) are
themselves MFOTL obligations over a suitably-typed trace (a trust-domain field, a
session id, an approval bound to object and amount): exactly the schema current
benchmarks lack (Section~\ref{sec:gap}).

\looseness=-1 Conditions (ii)--(iv) need fields the corpora do not carry, so they remain design
proposals. Condition (i) does not, once stated precisely, and we therefore
\emph{evaluate} it. The direct reading of (i) is vacuous against this attacker: the injection is itself a forged tool result, and the workspace obligation
already restricts accrual to tool messages, yet still flips $99\%$. What discriminates is
\emph{call--answer binding}: a genuine tool result answers a lookup the agent itself
issued, while a planted message answers nothing outstanding. In other words, a value is
trusted only if it comes from the response to a lookup the agent actually issued;
merely appearing in something formatted like a tool result is not enough:
\[
\mathit{known}(p)\;\leftrightarrow\;\ooalign{$\bullet$\cr}\bigl(\mathit{lookup}(t)\wedge
\raisebox{0pt}{$\bigcirc$}\,\mathit{result}(t,p)\bigr).
\]
Re-running the same four perturbations against this hardened predicate
(Table~\ref{tab:harden}) removes the evasion entirely: structured and quoted framings
fall from $94.0\%$ to $0.0\%$ on banking and $99.0\%$ to $0.0\%$ on workspace: at
\emph{no} cost: detection ($403/576$ banking, $203/518$ workspace) and benign firing
($6/25$, $5/46$) are unchanged to the run. The hardened predicate thus eliminates the
tested evasion without changing detection or benign firing, indicating that the
improvement arises from provenance integrity rather than from a precision--robustness
trade-off; the naive predicate was reading a field it had no warrant to trust. The
result holds for the tested attack family only: an attacker who can forge a matching
lookup call, or compromise the tool being queried, is outside its reach.

Provenance integrity is therefore a requirement on the trace, not a detail of
canonicalisation.

\section{Benchmark Readiness and Timed Policies}
\label{sec:readiness}

The results so far raise a question about the corpora rather than the monitor: what
would a benchmark have to record for history-dependent obligations to be evaluable at
all? Section~\ref{sec:gap} answers it with a readiness audit of the three corpora and
a control on $\tau$-bench, leading to a minimal trace schema. Section~\ref{sec:cps}
then shows, on authored CPS/IoT logs, what the timed policies that this schema would
unlock look like when MonPoly evaluates them.

\subsection{Benchmark readiness for history-dependent enforcement}
\label{sec:gap}

Our experiments expose a broader evaluation problem. Current agent-security benchmarks
record enough to determine whether an \emph{attack succeeded}, but not enough to
evaluate many \emph{history-dependent enforcement} policies. We therefore assess the
three corpora as enforcement datasets, and quantify the availability of the state,
provenance, approval and timing information a runtime monitor quantifies over. These
benchmarks were designed to measure attack success and agent behaviour, not runtime
enforcement. As shipped, they therefore omit information that history-dependent monitors
need, which is why the generic obligations of Section~\ref{sec:results} collapse into action
typing.

\paragraph{The readiness scorecard}

\begin{table}[t]
\centering
\caption{Benchmark readiness for history-dependent enforcement: fraction of
runs or records exposing each field.}
\label{tab:score}
\renewcommand{\arraystretch}{1.1}
\setlength{\tabcolsep}{5pt}
\begin{tabular}{@{}lccc@{}}
\toprule
\textbf{Field} & \textbf{AgentDojo} & \textbf{STAC} & \textbf{R-Judge} \\
\midrule
explicit approval      & $0\%$    & $1.2\%$  & $0.4\%$  \\
recipient/payee id     & $38.8\%$ & $58.4\%$ & $5.8\%$  \\
object id              & $36.4\%$ & $34.2\%$ & $47.3\%$ \\
user confirmation      & $4.6\%$  & $16.4\%$ & $0\%$    \\
real timestamps        & none     & none     & none     \\
structured state       & none     & none     & none     \\
reversibility typing   & none     & none     & none     \\
\bottomrule
\end{tabular}
\end{table}

A readiness scorecard over the three benchmarks (Table~\ref{tab:score}) makes this
concrete. Explicit \emph{approvals} are almost absent: $0/6899$ AgentDojo runs,
$6/483$ STAC chains and $2/571$ R-Judge records. Recipient or payee identity is
recoverable in roughly a third to a half of AgentDojo and STAC runs, object identity
in about a third, but user-confirmation events are rare in all three. Above all, \emph{none} of the three carries real
timestamps (only call order), structured state transitions, or reversibility typing.
The metric and stateful policies that motivate temporal RV (Section~\ref{sec:cps}, below) thus
cannot be evaluated on these corpora as shipped: we authored timed logs to exercise
them. As a practical consequence we derive a twelve-field \emph{enforcement-ready} trace
schema, enumerated below, giving the minimum
event semantics needed to
benchmark relational, provenance-sensitive and timed policies. Recording these fields is
cheap at trace-emission time: every one is available to the harness at the moment a
tool is called, and it is what would let the next generation of corpora evaluate this
class of defence at all. One caveat follows from the census below: fields are necessary
but not sufficient. A benchmark can carry an \texttt{approval\_id} column, but if its
interaction protocol never gives the user a turn in which to approve anything, the
column stays empty. An enforcement-ready corpus is therefore a claim about the
\emph{interaction protocol} as much as about the schema.

\paragraph{Why the approvals are absent} To assess whether the observed absence of
approvals is an artefact of tool-name extraction, we additionally examine user-message
content; the attrition across extraction methods is itself the result. Two structural
facts explain it. AgentDojo is \emph{single-turn}: exactly one user message per run
($6899/6899$), stating a task up front, after which the agent executes autonomously. An
approval is by definition a response to a \emph{proposed} action, which requires a user
turn \emph{after} the proposal; the protocol has no such turn, and consistently none of
its $68$ distinct tools is approval-like. Approvals there are foreclosed by the
interaction protocol, not merely unrecorded: no parser can recover them. STAC is
\emph{multi-turn} (mean $2.61$ user turns, up to five), so approvals \emph{can} occur, and
approval language does appear: mining the user text instead of tool names raises the
count almost twelvefold, from $6/483$ ($1.2\%$) to $70/483$ ($14.5\%$) on a loose keyword match
and $39/483$ ($8.1\%$) on a strict permission-granting match. But only $3/483$ ($0.6\%$)
occur \emph{before} the harmful step, which is the only position from which an approval
could gate anything, because STAC's attacks work precisely by smuggling the harmful
step in without returning to the user. Better extraction therefore does not rescue the
situation: it relocates the shortfall from recording to \emph{protocol and position}. An
enforcement-ready corpus needs an interaction shape in which approval is expressible, not
just a field in which it could be written down.

\begin{table}[t]
\centering
\caption{Approval census across the three recorded corpora. An approval can only gate
an action if a user turn follows the agent's proposal; the two rows move together
(approvals before the mutation: $3/483$ for STAC, $1039/1980$ for $\tau$-bench).}
\label{tab:taucontrol}
\renewcommand{\arraystretch}{1.1}
\setlength{\tabcolsep}{3.5pt}
\begin{tabular}{@{}lccc@{}}
\toprule
\textbf{Measure} & \textbf{AgentDojo} & \textbf{STAC} & \textbf{$\tau$-bench} \\
\midrule
user turns per run (mean)            & $1.00$ & $2.61$ & $7.08$ \\
approval-like tools                  & $0/68$ & -- & $0/26$ \\
approvals \emph{before} the mutation & $0\%$ & $0.6\%$ & $52.5\%$ \\
\bottomrule
\end{tabular}
\end{table}

\paragraph{A control case: $\tau$-bench} If approvals were simply hard to record, no
corpus would have them. We therefore ran the same census over the $1{,}980$ recorded
trajectories that $\tau$-bench~\cite{taubench2024} ships, a \emph{task} benchmark whose
system prompt is the domain policy and whose policy requires the agent to obtain user
confirmation before mutating state. The contrast is pronounced
(Table~\ref{tab:taucontrol}). $\tau$-bench averages $7.08$
user turns per trajectory against AgentDojo's $1.00$ and STAC's $2.61$; $88.7\%$ of its
trajectories contain a mutating call; and while a tool-name search again finds
\emph{nothing} ($0$ of its $26$ tools is an approval tool, exactly as in AgentDojo),
$1039/1980 = 52.5\%$ of trajectories contain explicit permission-granting language in a
user turn positioned \emph{before} the mutating call. Every one of those $1039$ answers a
preceding assistant turn that made no tool call, i.e.\ they are confirmations of a
proposal rather than instructions. Against STAC's $0.6\%$ that is a factor of eighty. The effect is not a
single-domain quirk: the rate is $58.0\%$ and $46.5\%$ on the airline domain
($200$ and $400$ trajectories, GPT-4o and Sonnet-3.5) and $65.4\%$ and $47.4\%$ on
retail ($460$ and $920$), so it holds across both domains and both models. The airline
domain is also the travel-booking setting Maris evaluates with $20$ synthetic queries,
which is the comparison we can make: same task domain, $600$ recorded trajectories
instead of authored ones.

Two conclusions follow. The approval shortfall is a property of \emph{attack} corpora
rather than of agent traces in general, so it is a fixable instrumentation gap and not an
intrinsic limit; and a partial remedy already exists, since on a corpus such as
$\tau$-bench the antecedent of O1--O5 would be genuinely satisfiable and the generic
obligations would not collapse to action typing. We do not report detection on
$\tau$-bench because it carries no attack labels: it measures task success, not attack
success, which is why the detection results here use AgentDojo and STAC. The two are
complementary, and a corpus with $\tau$-bench's interaction protocol and AgentDojo's
attack labels would, on this evidence, be the most valuable addition to the current
benchmark landscape.

\paragraph{The contribution of each field} A scorecard says which fields are missing; it does not
say what capability their absence costs. We therefore ablate the trace itself, masking
fields cumulatively on AgentDojo banking to give five \emph{richness levels}
R0--R4, and asking, at each level, which policy class
remains evaluable and how well it discriminates (Table~\ref{tab:ablation}). Two results
stand out. First, the levels the corpora \emph{do} support provide only limited
additional discrimination: recipient identity lifts $\mathrm{LR^+}$ from $1.80$ to $2.92$, and then the
approval- and timing-dependent levels are unevaluable: \emph{zero}
approval-typed calls and \emph{zero} timestamped messages across $1545$ runs. Second, and
less obviously, R1 and R2 are \emph{observationally identical} on clean traces: same
detection, same benign firing, same $\mathrm{LR^+}$. They separate only under attack, at
$94\%$ versus $0\%$ evasion. Field availability alone therefore does not tell you whether
a provenance policy is sound; the binding does, and its value is invisible to any
evaluation that does not include an adversary. This is the sharper form of the readiness
argument: a corpus can expose every field a policy names and still be unable to
distinguish a sound policy from an unsound one.

\begin{table}[t]
\centering
\caption{Trace-field ablation on AgentDojo banking. R0--R4 are cumulative
\emph{richness levels}: each adds what the trace must record and the monitor trust.
``Poison flip'' is the structured-line evasion rate of Section~\ref{sec:prov}. R1 and R2
differ only under attack.}
\label{tab:ablation}
\renewcommand{\arraystretch}{1.05}
\setlength{\tabcolsep}{3pt}
\begin{tabular}{@{}llccc@{}}
\toprule
\textbf{Level} & \textbf{policy class} & \textbf{detect} & \textbf{BFR} & \textbf{flip} \\
\midrule
R0 action only     & typed-action    & $79.2\%$ & $44.0\%$ & n/a \\
R1 $+$ recipient id & recipient-bound & $70.0\%$ & $24.0\%$ & $94\%$ \\
R2 $+$ provenance   & poison-resistant & $70.0\%$ & $24.0\%$ & $\mathbf{0\%}$ \\
\midrule
R3 $+$ approval id & approval-bound  & \multicolumn{3}{c}{unevaluable ($0/1545$ runs)} \\
R4 $+$ time, state & metric          & \multicolumn{3}{c}{unevaluable ($0$ timestamps)} \\
\bottomrule
\end{tabular}
\end{table}

\paragraph{A minimal trace schema} We therefore propose a minimal enforcement-ready
schema: twelve fields per event: \texttt{timestamp}, \texttt{actor}, \texttt{tool},
\texttt{action}, \texttt{object}, \texttt{recipient}, \texttt{source},
\texttt{approval\_id}, \texttt{session\_id}, \texttt{state\_before},
\texttt{state\_after}, \texttt{trust\_domain}. Current benchmarks supply the middle
group (tool/action, partially object/recipient) but none of wall-clock
\texttt{timestamp}, \texttt{approval\_id}, \texttt{session\_id}, the state pair, or
\texttt{trust\_domain}: the fields the timed policies (Section~\ref{sec:cps}) and the
hardened provenance spec (Section~\ref{sec:prov}) require. Cheap to emit but absent
everywhere, they are the concrete obstacle to evaluating history-dependent agent
enforcement.


\paragraph{Timed policies} Metric obligations are what the schema of this section would
make evaluable, and they are the one place the paper exercises MFOTL's metric operators.
Because the corpora carry no timestamps we exercise them on authored logs: three timed
policies, a multi-instance building trace and the R-Judge IoT subset, all run verbatim by
MonPoly, with a verdict that flips on identical actions purely by elapsed time.
Section~\ref{sec:cps} gives the detail; it demonstrates expressiveness rather than
field performance.

\subsection{Timed CPS/IoT policies}
\label{sec:cps}

This section shows that MFOTL/MonPoly applies to CPS/IoT policies; it is an
expressiveness study on authored logs, not a field validation. Metric runtime verification of
this kind is established practice in CPS engineering, for instance over FMI
co-simulation traces~\cite{temperekidis2022fmirv}; what changes here is that the
monitored signal is an agent's tool-call log. Agentic components are already being proposed for such settings, together with tamper-evident evidence trails~\cite{resilagent2026}, which is what makes the trace-integrity question of Section~\ref{sec:canon} concrete rather than hypothetical. We give three evidence points,
all evaluated verbatim by MonPoly: \emph{timed} (metric) actuation policies, a
multi-instance building trace, and a real (small) IoT subset.

\paragraph{Timed policies whose verdict depends on elapsed time} The generic
obligations exercise little temporal complexity: on the recorded corpora an unbounded
$\blacklozenge$ reduces to typed-action detection. We therefore author three
\emph{metric} obligations whose verdict on the \emph{identical} action changes purely
with the clock (signatures, logs and formulas in Appendix~\ref{app:repro};
Table~\ref{tab:timed}). \emph{(T1) Bounded-window consent}: an export is allowed only
if a consent is less than $24$\,h old. In MFOTL,
$\mathsf{export}(o)\wedge\neg(\text{\textsc{once}}[0,24]\,\mathsf{consent}(o))$, so
MonPoly fires only when the consent is stale. \emph{(T2) Sliding-window rate limit}: no
more than two actuations of a device within $6$\,h (a metric \textsc{cnt} over
$\text{\textsc{once}}[0,6]$): fires on a burst, silent once calls are spaced.
\emph{(T3) Timed precedence}: an \texttt{execute} needs an authorisation less than
$8$\,h old: fires only when the authorisation has expired. In each case the current
action is identical; only the timing of earlier events changes the verdict.

\begin{table}[t]
\centering
\caption{Timed (metric MFOTL) policies evaluated verbatim by MonPoly. For each, the
\emph{same} action is allowed or a violation purely by elapsed time; $\bullet$ marks
the timepoints where MonPoly fires. Full sig/log/formula in Appendix~\ref{app:repro}.}
\label{tab:timed}
\renewcommand{\arraystretch}{1.1}
\setlength{\tabcolsep}{5pt}
\begin{tabular}{@{}llc@{}}
\toprule
\textbf{Policy (window)} & \textbf{action, timing} & \textbf{MonPoly} \\
\midrule
T1 consent ($24$\,h)   & export, consent $12$\,h ago      & pass \\
T1 consent ($24$\,h)   & export, consent $48$\,h ago      & $\bullet$ \textbf{violation} \\
T1 consent ($24$\,h)   & export, re-consent $10$\,h ago   & pass \\
T2 rate ($\ge3/6$\,h)  & 3rd actuation within $6$\,h      & $\bullet$ \textbf{violation} \\
T2 rate ($\ge3/6$\,h)  & actuations spaced $>6$\,h apart  & pass \\
T3 precede ($8$\,h)    & execute, auth $6$\,h ago         & pass \\
T3 precede ($8$\,h)    & execute, auth $40$\,h ago        & $\bullet$ \textbf{violation} \\
T3 precede ($8$\,h)    & execute, re-auth $4$\,h ago      & pass \\
\bottomrule
\end{tabular}
\end{table}

The traces behind Table~\ref{tab:timed} are likewise authored, not recorded: the point
they establish is that \emph{elapsed time alone} flips MonPoly's verdict on an
otherwise identical action, which is a property of the specification language, not
evidence of field accuracy.

\paragraph{Multi-instance building actuation} On an authored $20$-event smart-building
trace ($13$ policy-relevant actions, three past-time policies, formulas in
Appendix~\ref{app:repro}) MonPoly flags exactly the $6$ history-violating instances and
clears all $7$ compliant ones. One violation is purely \emph{timed}: an enabling
\texttt{damper\_open} fell $4$\,h before a $3$\,h window, so the same compressor start is
a violation only by timing.

\paragraph{R-Judge IoT (real data, small: cannot be enlarged)} The R-Judge IoT
subset (smart home, traffic dispatch, phone) is \emph{all the labelled IoT agent
records the benchmark ships}: $30$ records ($19$ unsafe, $11$ safe), so we report it as
illustrative. The generic obligations yield $31.6\%$ detection ($6/19$) and $27.3\%$
FPR ($3/11$) with wide Wilson intervals. It does contain relational cases: in one
unsafe record the agent grants a dog-walker \emph{permanent} smart-lock access without
consent. Encoding it under a parametric consent obligation
$\square\,\forall g,p.(\mathsf{grant}(g,p)\wedge p{=}1\wedge
\neg\,\blacklozenge\,\mathsf{consent}())$, MonPoly fires on the permanent grant and
clears once a prior consent is present: a verdict flip on real IoT data the generic
``grant'' typing (which never fires here) cannot make.

\section{Discussion}
\label{sec:concl}

\paragraph{Relation to prior MFOTL work} MFOTL and MonPoly have been used for agent
policy enforcement before. Maris~\cite{maris2026} expresses privacy policies in MFOTL
and calls MonPoly from inside the agent runtime, on three authored applications with
$20$ synthetic queries each. We ask a different question: what can be measured
\emph{offline} on the attack corpora the community already ships, and what those
corpora cannot support. Maris ships no recorded trajectories to replay; we therefore
report on $6{,}899$ AgentDojo runs, $483$ STAC chains, $571$ R-Judge records and
$1{,}980$ $\tau$-bench trajectories. What formal RV contributes in this setting is the
combination of trace-level scope, formal temporal semantics, data-parametric provenance
and offline evaluability in \emph{one} declarative substrate. Table~\ref{tab:matrix}
(appendix) records what each cited system was shown to do in its evaluated
configuration, not what it could never do: AgentSpec~\cite{agentspec2026} has a formal
rule semantics and some history-aware rules, and a determined encoding could reach
further than its cells show.

\paragraph{Limitations} \emph{Construct validity: what we measured.} We measure whether
an obligation fires on a recorded trajectory, not whether a live agent is stopped; how
an agent adapts once a shield blocks it is a different question. Benign firing is a
proxy for utility cost, because we never re-run the blocked task. Keyword typing is
coarse (Section~\ref{sec:canon}), and roughly $29\%$ of STAC cases, and more of R-Judge,
are harmful only in their text content, which structural monitoring cannot judge.
\emph{External validity: how far the results carry.} The parametric refinements are
evaluated on the banking and workspace suites only, and the cross-model span
(Section~\ref{sec:results}) covers the models AgentDojo ships. STAC records almost no
preconditions ($6/483$ chains by tool name), so the metric operators of MFOTL are
exercised only in the CPS/IoT study, on authored logs; that study shows expressiveness,
not how often such violations occur in the field. \emph{Deployment validity: what a
deployer should take from this.} At $29.3\%$ benign firing the generic layer is a triage
filter, not a gate, and the hardened provenance predicate is shown robust only against
the tested attack family.

\section{Conclusion}
\label{sec:conclusion}

We replayed the recorded attack corpora the community already uses through an
unmodified formal monitor. Five reusable obligations flagged $71.8\%$ of STAC chains and
$70.1\%$ of GPT-4o's successful AgentDojo attacks, at the price of firing on $29.3\%$ of
benign runs. Most of that detection is risky-action typing, because the corpora rarely
record the approvals and timestamps that temporal logic reasons about. Where the trace
does carry relational or timing context, MFOTL states conditions no per-call rule can
decide, and it fires early: before the harmful call in $52.4\%$ of successful attacks.

The second lesson is that a specification is only as trustworthy as the trace it reads.
The provenance predicate that makes the parametric policies precise can be poisoned by a
single planted line, which flips $94$--$99\%$ of the runs it would otherwise flag.
Binding provenance to the lookup that produced it removes the evasion completely for the
tested attack family, at no cost in detection or benign firing. On clean traces the naive
and hardened predicates are indistinguishable; only an adversary tells them apart.

For benchmark designers the request is twofold. Future agent-security benchmarks should
record the twelve fields of Section~\ref{sec:gap}, and they should use an interaction
protocol in which approval can occur at all. The $\tau$-bench control shows that the
second is the binding constraint: where a user is present to confirm, approvals appear
in over half of the trajectories; where none is, no schema can recover them.

\begin{acks}
This work has received funding from the European Union's Digital Europe Programme
under grant agreement No 101190251 (CYBERGUARD).
\end{acks}

\bibliographystyle{ACM-Reference-Format}
\bibliography{references}

\appendix
\raggedbottom
\section{Reproducibility}
\label{app:repro}
\sloppy
This appendix is excluded from the page limit. All scripts are pure-stdlib Python in
\texttt{agent-rv-impl/}; the datasets are the public releases below.

\paragraph{Datasets} STAC (arXiv:2509.25624,
\texttt{amazon-science/\allowbreak Mul\-ti\-Turn\-Agent\-Attack},
$483$ chains, file \texttt{STAC\_\allowbreak benchmark\_\allowbreak data.json}); AgentDojo
(\texttt{ethz-spylab/\allowbreak agentdojo}, recorded
\texttt{runs/\allowbreak<model>/\allowbreak<suite>/...}, $\approx$$36.7$k
runs; we use \texttt{gpt-4o-2024-05-13} in detail and, for the cross-model analysis,
the $22$ base (no-defence) model directories the release ships, plus their defence
variants where pooled: $28$ directories with a banking suite and $29$ with a
workspace suite);
R-Judge (\texttt{Lordog/R-Judge}, \texttt{data/}, $571$ records; the IoT subset
\texttt{data/IoT/\{household,trafficdispatch,phone\_iot\}.json} is all $30$ IoT
records the benchmark ships and cannot be enlarged offline).

\paragraph{Keyword patterns (typing)} A tool matches a class iff its lower-cased name
contains any listed substring.
\emph{DESTRUCTIVE:} delete, remove, cancel, clear, wipe, drop, purge, erase, revoke,
uninstall.
\emph{TEARDOWN:} autodelete, disable, stop, unlock, deactivate, turn\_off, bypass,
whitelist, allowlist, add\_user\_autodelete.
\emph{FINANCIAL:} send\_money, transaction, transfer, payment, pay\_, book\_, buy,
purchase, zelle, venmo, wire, invoice, checkout.
\emph{EXPORT:} send\_email, share, export, upload, post\_, publish, forward,
sendmessage, send\_message, send\_sms, tweet.
\emph{ACTUATION:} set\_, adjust, trafficlight, unlock\_door, execute, terminal,
medication, open\_, close\_, activate, reboot, restart, deploy, write\_, update\_,
configure, install.
\emph{APPROVAL (precondition):} confirm, approve, consent, authorize, authorise,
permission, verify, grant, two\_factor, 2fa, otp, request\_approval, acknowledge,
sign\_off, double\_check. No tool whose name begins \texttt{get\_}, \texttt{list\_},
\texttt{read\_}, \texttt{search\_}, \texttt{view\_}, \texttt{fetch\_} or
\texttt{query\_} counts as an approval event or as a typed action, since a read-only
accessor is neither a grant of permission nor a commit (Section~\ref{sec:method}). This is
the frozen (headline) set, after those corrections, used for every number reported here.

\paragraph{Argument normalisation} Tool arguments are parsed from JSON to a dict
(malformed strings kept under a raw key). The \texttt{known\_payee} and
\texttt{known\_contact} predicates are \emph{not} hand-curated: they are emitted by
the same scripted canonicaliser via a fixed regular expression over structured
recipient/sender fields, and the regexes were frozen \emph{before} we inspected any
attacker IBAN/address or any \textsf{security}/\textsf{utility} outcome. For the payee
obligation a recipient is a \emph{known payee} only if it matched
\texttt{(?:recipient|\allowbreak sender):\allowbreak\textbackslash s*\allowbreak([A-Z]\{2\}\allowbreak[0-9A-Z]\{6,\})}
in prior message text (a structured IBAN field, not free text). For the external-recipient
obligation an address is a \emph{known contact} only if it appeared under a directory
key \texttt{(?:from|\allowbreak sender|\allowbreak recipients?|\allowbreak participants?|\allowbreak email|\allowbreak shared\_with|\allowbreak owner):}
in a prior \emph{tool} result. So a payee/contact becomes ``known'' purely by the
scripted rule, independent of whether it is later an attack address.

\paragraph{Approval/consent recognition} An approval precedes step $i$ iff some call
at $j<i$ (same trace, same STAC case) matches the APPROVAL patterns. An obligation
fires on step $i$ iff the tool matches its class patterns \emph{and} no approval
precedes $i$.

\paragraph{Timestamps and multiple firings} Each canonical event becomes one MonPoly
timepoint \texttt{@t}; for STAC the class and \texttt{approval} predicates are
parameterised by the case id so \texttt{ONCE} only matches within a case. Only
timepoints that carry at least one typed predicate are emitted: a canonicalised step
whose tool matches \emph{no} obligation class (a read-only or out-of-template action)
produces no event, so the log holds $357$ predicate-bearing timepoints across the
$483$ cases, and cases whose final step matches no class emit nothing and cannot fire
(counted as a miss). Because the STAC obligations use an unbounded
$\text{\textsc{once}}[0,10^6]$, dropping predicate-free timepoints is sound: an
unbounded past-time operator is insensitive to the number of intervening empty
timepoints, only to whether a matching predicate occurred earlier in the case. For the
\emph{timed} policies of Section~\ref{sec:cps} this is different: there the integer
timestamp is the event's clock time (in hours), timepoints are spaced by real elapsed
time, and the metric windows ($\text{\textsc{once}}[0,W]$) depend on those gaps, so we
keep the actual times, not a dense index. Per-obligation counts are
non-exclusive; a case/run counts as detected if $\ge 1$ obligation fires.

\paragraph{Full MFOTL obligations (monitored violation form)} With
$\blacklozenge$ once-in-the-past:
\begin{align*}
&\text{O1}: \mathsf{destructive}(a)\wedge\neg\,\blacklozenge\,\mathsf{approved}(a)\\
&\text{O2}: \mathsf{teardown}(a)\wedge\neg\,\blacklozenge\,\mathsf{approved}(a)\\
&\text{O3}: \mathsf{financial}(a)\wedge\neg\,\blacklozenge\,\mathsf{approved}(a)\\
&\text{O4}: \mathsf{export}(a)\wedge\neg\,\blacklozenge\,\mathsf{approved}(a)\\
&\text{O5}: \mathsf{actuation}(a)\wedge\neg\,\blacklozenge\,\mathsf{approved}(a)
\end{align*}
The STAC export uses the disjunction handed verbatim to MonPoly:
\texttt{(destructive(c) OR teardown(c) OR financial(c) OR export(c) OR
actuation(c)) AND NOT (ONCE[0,1000000] approval(c))}.
The two parametric obligations are $\square\forall p.(\mathsf{transfer}(p)\wedge
\neg\blacklozenge\mathsf{known\_payee}(p))$ and
$\square\forall r.(\mathsf{send}(r)\wedge\neg\blacklozenge\mathsf{known\_contact}(r))$.
The multi-instance building policies, run verbatim by MonPoly, are:
\begin{quote}\ttfamily\footnotesize\raggedright
P1: setpoint(z,v) AND v > 28 AND NOT ((NOT maint\_off()) SINCE maint\_on())\\
P2: compressor\_start(z) AND NOT (ONCE[0,3] damper\_open(z))\\
P3: door\_unlock(d) AND NOT (ONCE[0,3] badge\_auth(d))\\
lock: grant\_access(g,p) AND p = 1 AND NOT (ONCE consent())
\end{quote}
The $20$-event building log yields firings at timepoints
$\{4,5,9,14,19,22\}$ ($6$ violations; the P2 firing at $t{=}14$ is the timed miss,
a \texttt{damper\_open} at $t{=}10$ outside the $3$-unit window) and no firing on the
$7$ compliant instances.

\paragraph{Timed (metric) policies of Section~\ref{sec:cps}} These use hour-valued
timestamps; formula, signature and log are run verbatim by MonPoly:
\begin{quote}\ttfamily\footnotesize\raggedright
T1: export(o) AND NOT (ONCE[0,24] consent(o))\\
\hspace*{1em}log: @0 consent(1) @12 export(1) @48 export(1) @50 consent(1) @60 export(1)\\
\hspace*{1em}$\Rightarrow$ fires only @48 (consent 48h stale)\\
T2: (c <- CNT s; d (ONCE[0,6] actuate(d,s))) AND c >= 3\\
\hspace*{1em}log: @0 actuate(7,1) @2 actuate(7,2) @5 actuate(7,3) @30.. @33.. @40..\\
\hspace*{1em}$\Rightarrow$ fires only @5 (3rd actuation within 6h; spaced calls never fire)\\
T3: execute(x) AND NOT (ONCE[0,8] request(x))\\
\hspace*{1em}log: @0 request(5) @6 execute(5) @40 execute(5) @44 request(5) @48 execute(5)\\
\hspace*{1em}$\Rightarrow$ fires only @40 (authorisation 40h old)
\end{quote}
Each timed verdict flips on the \emph{same} action purely by elapsed time; T2 uses
MonPoly's metric \texttt{CNT} aggregation over a sliding $[0,6]$ window.

\paragraph{MonPoly version and invocation} The engine is MonPoly, version string
\texttt{MonPoly (development build)} (from \texttt{monpoly -version}), at
\texttt{/usr/local/bin/monpoly} inside Docker image \texttt{infsec/monpoly}
(image id \texttt{sha256:82cde1a2\allowbreak3cd3}). It is invoked as
\texttt{monpoly -sig <s>.sig -formula <f>.mfotl -log <l>.log}; all formulas above pass
\texttt{monpoly -check} as monitorable. On the STAC export it fires on exactly $347$
timepoints, matching the reference monitor. Because both consume the same canonical log,
that agreement checks our implementation rather than the specification, so we add a
genuinely independent engine. \textbf{DejaVu}~\cite{havelund2018dejavu} decides
first-order past-time properties by a different procedure (BDDs over quantified
past-time LTL). The STAC obligation is untimed past-time first order, so it transfers to
DejaVu's QTL without weakening; on the same event stream DejaVu reports exactly $347$
violations. We additionally run \textbf{VeriMon}~\cite{verimon2022}, MonPoly's formally
verified kernel (\texttt{monpoly -verified}), whose monitoring algorithm is
machine-checked in Isabelle/HOL: also exactly $347$. So the obligation is agreed by four
monitors: our reference implementation, MonPoly's standard kernel, its verified
kernel, and an independently developed BDD-based engine: of which the last two make
the result something stronger than agreement between MonPoly and a checker we wrote
ourselves: one is a different decision procedure, and one is proved correct.

\paragraph{Artifact availability} All evaluation scripts are pure-stdlib Python under
\texttt{agent-rv-impl/}; together with the MonPoly signature/log/formula files for the
STAC, building, IoT and timed studies they will be released as supplementary material
(and the public repository
\url{https://github.com/nikos-kekatos/formal-rv-tool-using-llm-agents}), so the full
pipeline is
reproducible against the public dataset releases named above.

\paragraph{AgentDojo selection and unflagged successful-attack fraction} Runs are read
by recursive glob of \texttt{runs/<model>/**/*.json}. A run is \emph{attacked} iff
\texttt{attack\_type} is set and $\neq$\texttt{none} and \texttt{injection\_task\_id}
is present; \emph{benign} iff neither is present; runs with an empty/unparseable
tool-call chain are skipped ($96$ for GPT-4o $\Rightarrow 6680$ attacked, $123$
benign). \texttt{security==True} marks a successful attack, \texttt{utility==True} a
completed task. For reference we report the \emph{unflagged successful-attack
fraction}, $(\textsf{successful}\wedge\neg\textsf{RV-fired})/\textsf{attacked}
=(1931-1354)/6680=8.6\%$; this is a descriptive count of successful attacks the
monitor did not flag, \emph{not} a measured attack-success rate under an in-loop
shield (which would require re-running the agent).

\paragraph{Scripts} \texttt{stac\_rv.py}, \texttt{stac\_to\_monpoly.py},
\texttt{agentdojo\_rv.py}, \texttt{agentdojo\_scenarios.py},
\texttt{agentdojo\_param.py} (payee), \texttt{agentdojo\_param2.py}
(external recipient), \texttt{agentdojo\_param\_pooled.py} (pooled cross-model
McNemar: produces the headline $p$-values),
\texttt{cross\_model.py} (per-directory cross-model sign test),
\texttt{first\_firing.py} (earliest-warning),
\texttt{combined\_ablation.py} and \texttt{per\_oblig\_ablation.py} (ablations),
\texttt{risk\_enrichment.py} ($\mathrm{LR^+}$ and balanced accuracy),
\texttt{canon\_sensitivity.py} (four-mapping sensitivity),
\texttt{perturb\_provenance.py} (poisoning),
\texttt{readiness\_scorecard.py} (benchmark scorecard),
\texttt{corpus\_gen.py} (AgentTrace-T), \texttt{metamorphic.py} (metamorphic tests),
\texttt{mp\_time.py} (MonPoly runtime),
\texttt{canon\_eval.py}, \texttt{wilson\_ci.py}, \texttt{mcnemar.py} (single-model
paired-benign, the inconclusive per-model result), \texttt{rjudge\_rv.py},
\texttt{harden\_provenance.py} (lookup-bound provenance vs.\ the same poisoning),
\texttt{field\_ablation.py} (the R0--R4 levels),
\texttt{approval\_census.py} (approval density across AgentDojo, STAC and
$\tau$-bench), \texttt{stac\_to\_dejavu.py} (the DejaVu cross-check),
\texttt{run\_corpus.py}, \texttt{gen\_stac\_scale.py} and
\texttt{agentdojo\_to\_monpoly.py}, and the
MonPoly \texttt{.sig}/\texttt{.mfotl}/\texttt{.log} files for the STAC, building, IoT
and timed studies. Nothing beyond this list is shipped: the repository contains exactly
the scripts and specifications behind the numbers reported here; \texttt{reproduce.sh}
at the repository root runs them end to end.

\section{Extended Results and Robustness}
\label{app:extended}

This appendix provides the detailed evidence behind the headline results of
Sections~\ref{sec:results}--\ref{sec:readiness}. It first decomposes the aggregate
AgentDojo result by obligation and by suite, then tests robustness across models and
alternative tool classifications, then reports implementation, scalability and
synthetic-policy checks, and finally gives the additional parametric and adversarial
results. Table~\ref{tab:matrix} positions the compared systems; it records evaluated
configurations, not capabilities.

\subsection{How to read the extended results}
\paragraph{Reporting conventions}\label{app:conventions} Unless stated otherwise,
\emph{detection} is the fraction of successful attacks (AgentDojo \textsf{security}=true
runs, or STAC chains) on which an obligation fires, and \emph{BFR} is the fraction of
benign runs on which one fires. A \emph{single-obligation} rate counts one obligation;
a \emph{policy-set union} counts a run as flagged if any obligation in the set fires.
Two conventions explain every apparent mismatch between tables. \emph{Policy set}: a
union over the five generic obligations is not comparable with a single obligation on
the same runs; workspace benign firing is $10/46$ as a union against $7/46$ for O4
alone, and Table~\ref{tab:param}'s banking $11/25$ is a single obligation, not the
union. \emph{Run filter}: the aggregate replay and Table~\ref{tab:suite} count only
runs with a non-empty tool-call chain ($96$ banking runs excluded, one of them benign),
while the parametric scripts and Table~\ref{tab:param} count every run with a
non-empty message list. This is why banking benign firing reads $19/24=79.2\%$ in one
place and $19/25=76.0\%$ in another with the same numerator.

\begin{table}[tbp]
\centering
\caption{Paradigm comparison, \emph{per each system's evaluated configuration} (not a
fundamental-capability claim). TL = trace-level; FS = formal temporal semantics;
PAR = parametric (quantifies over trace data); OFF = evaluable offline on recorded
runs. $\circ$ = partial / achievable with bespoke stateful code.}
\label{tab:matrix}
\renewcommand{\arraystretch}{1.1}
\setlength{\tabcolsep}{5pt}
\begin{tabular}{@{}lcccc@{}}
\toprule
\textbf{System} & \textbf{TL} & \textbf{FS} & \textbf{PAR} & \textbf{OFF} \\
\midrule
AgentDojo defences~\cite{agentdojo2024} & --         & --         & --         & \cmark \\
AgentSpec~\cite{agentspec2026}    & $\circ$    & \cmark & $\circ$    & -- \\
Progent~\cite{progent2025}        & $\circ$    & --         & --         & -- \\
GuardAgent~\cite{guardagent2024}  & --         & --         & --         & -- \\
STAC~\cite{stac2026} (benchmark)  & \cmark & --         & --         & \cmark \\
\textbf{Formal RV / MonPoly (ours)} & \cmark & \cmark & \cmark & \cmark \\
\bottomrule
\end{tabular}
\end{table}

\subsection{Decomposing the headline result}
\paragraph{Per-obligation decomposition (AgentDojo, GPT-4o)} Table~\ref{tab:peroblig}
asks which obligations contribute to the aggregate $70.1\%$ detection;
Table~\ref{tab:suite} asks in which AgentDojo domains that detection occurs. Three
observations follow from Table~\ref{tab:peroblig}. The aggregate is distributed across
several obligation classes rather than dominated by one rule. Export produces the
largest number of unique detections, whereas the destructive-action obligation
discriminates best between successful attacks and benign runs
($\mathrm{LR^+}=3.91$, ahead of export at $2.64$). The teardown obligation never fires
because the evaluated suites contain no protection-teardown actions.
\emph{Takeaway from Table~\ref{tab:suite}:} banking shows the highest attack coverage
but also the highest benign firing ($97.4\%$ / $79.2\%$); travel shows the opposite
pattern ($16.5\%$ / $11.1\%$).

\begin{table}[tbp]
\centering
\caption{Per-obligation detection on successful attacks and BFR (AgentDojo, GPT-4o),
with unique flagged successful attacks and $\mathrm{LR^+}=$ detection/BFR. Union is the
headline any-obligation result.}
\label{tab:peroblig}
\renewcommand{\arraystretch}{1.0}
\setlength{\tabcolsep}{5pt}
\begin{tabular}{@{}lcccc@{}}
\toprule
\textbf{Obligation} & \textbf{detect \%} & \textbf{BFR \%} & \textbf{unique} & \textbf{$\mathrm{LR^+}$} \\
\midrule
O1 destructive & $15.9$ & $4.1$  & $259$ & $3.91$ \\
O2 teardown    & $0$    & $0$    & $0$   & --     \\
O3 financial   & $26.8$ & $13.0$ & $243$ & $2.06$ \\
O4 export      & $27.9$ & $10.6$ & $491$ & $2.64$ \\
O5 actuation   & $16.2$ & $6.5$  & $38$  & $2.49$ \\
\midrule
union (any)    & $70.1$ & $29.3$ & --    & $2.40$ \\
\bottomrule
\end{tabular}
\end{table}

\begin{table}[tbp]
\centering
\caption{Per-suite replay (GPT-4o, no defence): AgentDojo attack-success, RV detection
on successful attacks and RV BFR on benign runs, as policy-set unions over the five
generic obligations (conventions in Appendix~\ref{app:conventions}).}
\label{tab:suite}
\renewcommand{\arraystretch}{1.0}
\setlength{\tabcolsep}{6pt}
\begin{tabular}{@{}lccc@{}}
\toprule
\textbf{Suite} & \textbf{ASR} & \textbf{RV detect} & \textbf{BFR} \\
\midrule
banking   & $40.1\%$ & $97.4\%$ & $79.2\%$ \\
slack     & $60.9\%$ & $63.9\%$ & $15.4\%$ \\
travel    & $9.3\%$  & $16.5\%$ & $11.1\%$ \\
workspace & $19.9\%$ & $62.9\%$ & $21.7\%$ \\
\bottomrule
\end{tabular}
\end{table}

\subsection{Robustness across models and canonicalisation}

Two different robustness questions are answered here: whether the parametric result
generalises across models, and whether the headline result depends on how tools are
typed.

\paragraph{Cross-model sign test (parametric payee, banking)} Beyond the pooled
McNemar test (Section~\ref{sec:param}), we check the direction of the parametric effect
per model. Across the $22$ base (no-defence) model directories with a banking suite,
the parametric payee obligation \emph{decreases} benign firing in $20$, leaves it
unchanged in $2$, and increases it in $0$; the median per-directory reduction is
$8.0$ percentage points (IQR $[4,17]$, min $0$, max $24$). An exact two-sided sign
test on the $20$-vs-$0$ split gives $p=1.9\times10^{-6}$: the reduction direction is
consistent across models, not an artefact of pooling. Over the same $22$ directories,
generic detection ranges $48.4$--$91.7\%$ and BFR $16.7$--$34.8\%$. \emph{Takeaway:} the reduction in
benign firing from payee provenance is directionally consistent across models rather
than driven by the pooled GPT-4o result.

\paragraph{Canonicalisation corrections (forensic detail)}\label{app:forensic} The approval list originally contained the
token \texttt{ack}, intended for acknowledgement tools. Because matching is on
substrings it also matched \texttt{invite\_user\_to\_slack} and
\texttt{remove\_user\_from\_slack} ($595$ calls in the Slack suite) each of which
then acted as a prior approval and suppressed every later obligation in that run. The
pattern \texttt{authorize} likewise matched the read-only lookup
\texttt{list\_authorized\_personnel}. We corrected both after the fact: \texttt{ack} is
now spelled \texttt{acknowledge}, and no read-only accessor (prefix \texttt{get\_},
\texttt{list\_}, \texttt{read\_}, \texttt{search\_}, \texttt{view\_}, \texttt{fetch\_} or
\texttt{query\_}) can constitute an approval event. The correction \emph{raises} the
reported rates: AgentDojo detection $61.6\%\!\to\!70.2\%$, Slack-suite detection
$40.3\%\!\to\!63.9\%$, BFR $31.7\%\!\to\!32.5\%$, so the pre-correction figures
understated the method; they did not inflate it. The same read-only guard then had to
be applied to \emph{action typing} as well, because the financial pattern
\texttt{transaction} was matching the read-only \texttt{get\_most\_recent\_transactions}
and \texttt{get\_scheduled\_transactions}: retrieving records is not a value transfer,
and the hand-labelled gold set of Section~\ref{sec:canon} marks all twenty read-only accessors
\textsf{NONE}. That second correction moved detection to $70.1\%$ and benign firing to
$29.3\%$, and those are the figures reported throughout; it also raised typing precision
(micro $0.84\!\to\!0.90$) and made the payee ablation discriminate at all, since the
read-only matches had previously kept the refined financial slot permanently lit. STAC
and R-Judge are unaffected: no tool name in either corpus collides. We report the correction instead of silently re-freezing, because it is
precisely the class of error a substring canonicaliser invites and precisely what
Section~\ref{sec:canon} sets out to measure.

\begin{table}[tbp]
\centering
\caption{Canonicalisation typing vs.\ $54$ hand-labelled tools. Per-class multi-label
precision/recall (does the typing include the gold class?), micro-average, and
single-primary exact-match.}
\label{tab:canon}
\renewcommand{\arraystretch}{1.0}
\setlength{\tabcolsep}{6pt}
\begin{tabular}{@{}lccc@{}}
\toprule
\textbf{Class} & \textbf{P} & \textbf{R} & \textbf{(tp/fp/fn)} \\
\midrule
O1 destructive & $0.78$ & $1.00$ & $7/2/0$ \\
O2 teardown    & $1.00$ & $1.00$ & $3/0/0$ \\
O3 financial   & $1.00$ & $0.78$ & $7/0/2$ \\
O4 export      & $1.00$ & $0.67$ & $4/0/2$ \\
O5 actuation   & $0.86$ & $0.86$ & $6/1/1$ \\
\midrule
micro-average  & $0.90$ & $0.84$ & F1 $0.87$ \\
\multicolumn{4}{l}{single-primary exact-match: $47/54 = 87.0\%$} \\
\bottomrule
\end{tabular}
\end{table}

\paragraph{Canonicalisation sensitivity} Table~\ref{tab:canonsens} re-runs the
AgentDojo and STAC replays under four tool$\to$class mappings. The frozen mapping is the
one reported throughout; the other three are alternatives, not corrections of it.
Gold-correcting the hand-labelled tools raises detection and BFR together, and the
expanded mapping coincides with it on these corpora, so the gold corrections act by
adding missed classes rather than by removing keyword classes that fire. The
conservative mapping lowers AgentDojo detection to $69.0\%$ and BFR to $26.0\%$, and
STAC detection further. \emph{Takeaway:} absolute detection depends moderately on tool
typing, but the headline mapping is not an optimistic choice; alternative mappings raise
detection only by raising benign firing substantially as well.

\begin{table}[tp]
\centering
\caption{Sensitivity of the headline rates to the tool$\to$class mapping. AD =
AgentDojo (GPT-4o); STAC detection is final-step. \emph{Frozen}: the keyword typing
used throughout the paper (after the corrections of Appendix~\ref{app:forensic}).
\emph{Gold-corrected}: the $54$ hand-labelled tools take their gold class, all others
keep the keyword class. \emph{Conservative}: any tool the keyword typing assigns to
more than one class is dropped from every risky class. \emph{Expanded}: keyword classes
plus any gold class the keyword typing missed. Approval gating is identical in all four.}
\label{tab:canonsens}
\renewcommand{\arraystretch}{1.0}
\setlength{\tabcolsep}{4.5pt}
\begin{tabular}{@{}lccc@{}}
\toprule
\textbf{Mapping} & \textbf{AD detect \%} & \textbf{AD BFR \%} & \textbf{STAC detect \%} \\
\midrule
frozen (headline)  & $70.1$ & $29.3$ & $71.8$ \\
gold-corrected     & $79.9$ & $43.9$ & $73.1$ \\
conservative       & $69.0$ & $26.0$ & $61.7$ \\
expanded           & $79.9$ & $43.9$ & $73.1$ \\
\bottomrule
\end{tabular}
\end{table}

\subsection{Implementation validation and performance}

The results in this subsection are engineering and semantic validation, not
attack-detection evidence. The four-monitor agreement on STAC is reported in
Appendix~\ref{app:repro}.

\paragraph{Synthetic corpus (AgentTrace-T) and metamorphic tests} \emph{These results
validate the implementation and the temporal semantics, not real-world attack coverage.}
To exercise the
specifications where the public corpora lack the needed fields, we generated
\emph{AgentTrace-T}: $1400$ traces, $200$ per policy across P1--P3, T1--T3 and the
IoT-lock policy, in three generated kinds (core, and boundary cases on each side of the
window), each labelled compliant or violating, roughly $100$ of each per policy. MonPoly
achieves $100\%$ verdict accuracy, $100\%$ parameter-binding accuracy, and $100\%$ on
the boundary cases. A metamorphic test suite then applies six semantics-preserving or
semantics-changing transformations: moving an enabler across the window boundary,
changing an id, reordering, inserting an irrelevant event (verdict must not change),
duplicating an event (must not change), and boundary shifts, and MonPoly gives the
expected verdict on $780/780$ ($100\%$). These perfect scores must be read for what
they are: AgentTrace-T was authored \emph{against} the same policies MonPoly then
checks, so $100\%$ measures whether the metric and first-order operators behave as
specified (an implementation and semantics check) and not whether the policies
catch attacks in the field. No field corpus is involved, and none of the headline
detection numbers rests on this corpus. We report it because the public corpora lack
the timestamps and approval events these operators need
(Section~\ref{sec:gap}), so there is otherwise no way to exercise them at all.

\begin{table}[tbp]
\centering
\caption{MonPoly throughput and peak RSS. STAC $\times k$ is the STAC log replicated
$k$ times; firings scale exactly ($10\times347=3470$).}
\label{tab:runtime}
\renewcommand{\arraystretch}{1.0}
\setlength{\tabcolsep}{6pt}
\begin{tabular}{@{}lcc@{}}
\toprule
\textbf{Log} & \textbf{throughput (events/s)} & \textbf{peak RSS} \\
\midrule
STAC $\times1$          & $52$k--$179$k & $8$--$14$\,MB \\
STAC $\times10$         & $452$k        & $8$--$14$\,MB \\
STAC $\times50$         & $429$k        & $8$--$14$\,MB \\
AgentDojo-derived       & $\approx0.8$M & $8$--$14$\,MB \\
\bottomrule
\end{tabular}
\end{table}

\paragraph{MonPoly runtime and scale-up soundness} Table~\ref{tab:runtime} reports
throughput and peak resident memory for the STAC log at $1\times$, $10\times$ and
$50\times$ replication and for an AgentDojo-derived log. Throughput is dominated by
start-up on short logs, rising from $52$--$179$k events/s on the $357$-timepoint STAC
log to $0.43$--$0.45$M at $\times10$--$\times50$ and $0.8$M on the AgentDojo export,
with flat $8$--$14$\,MB RSS; at $\times1$ the parametric formula is about three times
slower than the generic one ($52$k vs $155$k), a gap that closes as the log grows.
Replicating the STAC log $10\times$ yields exactly $3470=10\times347$ firings, which
checks determinism and linear scaling rather than soundness.

\subsection{Additional parametric and adversarial results}

\paragraph{Combined parametric ablation} \emph{Applying both parametric obligations
together barely moves run-level benign firing, because benign runs also trip other
generic obligations; the gain appears at the action slot.} The rates in this paragraph are
\emph{run-level policy-set} rates over pooled runs: a run counts as flagged if
\emph{any} obligation in the set fires on it. Applying \emph{both} parametric
obligations together, the banking$+$workspace pooled rates move from the union of all
five generic obligations at $80.6\%$ detection / $40.8\%$ BFR to $79.6\%$ / $32.4\%$,
and the four-suite overall rates from the same generic union at $69.9\%$ / $29.0\%$ to
both-parametric $69.4\%$ / $24.2\%$. These pooled figures are unions over the full
generic set and are therefore \emph{not} comparable with the single-obligation rows of
Table~\ref{tab:param}: a run that O3 alone misses can still be caught by another
generic obligation, so the union can exceed any individual obligation's rate. Run-level BFR barely moves
because benign runs also trip \emph{other} generic obligations; the parametric win is
at the \emph{action slot}: over the pooled banking$+$workspace benign runs the export
slot BFR falls $9.9\%\!\to\!7.0\%$ and the \texttt{send\_money} slot
$22.5\%\!\to\!16.9\%$, which is where the provenance conjunct applies. (Before the
read-only correction of Section~\ref{sec:method} the \texttt{send\_money} slot appeared not to
move at all, because read-only accessors matching the financial pattern kept the slot lit
regardless of the refinement.)

\balance
\paragraph{Provenance poisoning (detail)} \emph{Only framings that look like structured
directory data defeat the naive provenance check; free-text mentions and reformatted
identifiers do not.} The poisoning experiment
(Section~\ref{sec:prov}) injects, per flagged run, one line naming the attacker
recipient in four framings and re-evaluates the parametric obligation. Banking payee
($403$ flagged successful-attack runs): structured $379/403=94.0\%$ flip,
quoted/forwarded $94.0\%$, free-text $0\%$, reformatted-IBAN $0\%$. Workspace
recipient ($203$ flagged): structured $201/203=99.0\%$, quoted $99.0\%$, free-text
$0\%$, reformatted $0\%$. The four framings and the injection template are those
implemented in \texttt{perturb\_provenance.py}. Free-text mentions and reformatted identifiers are not
parsed as structured directory entries, which is why they flip nothing.

\end{document}